\documentclass[12pt]{article}
\usepackage{amssymb,amsfonts}
\newlength{\extraspace}
\newlength{\extraspaces}
\newcommand{\be}{\begin{equation}
\addtolength{\abovedisplayskip}{\extraspaces}
\addtolength{\belowdisplayskip}{\extraspaces}
\addtolength{\abovedisplayshortskip}{\extraspace}
\addtolength{\belowdisplayshortskip}{\extraspace}}
\newcommand{\ee}{\end{equation}}

\newcommand{\ba}{\begin{eqnarray}
\addtolength{\abovedisplayskip}{\extraspaces}
\addtolength{\belowdisplayskip}{\extraspaces}
\addtolength{\abovedisplayshortskip}{\extraspace}
\addtolength{\belowdisplayshortskip}{\extraspace}}
\newcommand{\ea}{\end{eqnarray}}

\newcommand{\newsection}[1]{
\vspace{15mm}
\pagebreak[3]
\addtocounter{section}{1}
\setcounter{equation}{0}
\setcounter{subsection}{0}
\setcounter{footnote}{0}
\begin{flushleft}
{\large\bf \thesection. #1}
\end{flushleft}
\nopagebreak
\medskip
\nopagebreak}

\newcommand{\Tr}{{\rm Tr}}
\newcommand{\Ucr}{U^{\dagger}}
\newcommand{\Xcr}{X^{\dagger}}

\newcommand{\gru}{U(L)\otimes U(L)}
\newcommand{\La}{\mathcal{L}}
\newcommand{\sgru}{SU(L)\otimes SU(L)}
\newcommand{\I}{\textbf{I}}
\newcommand{\csti}{F^{2}_{\pi}+LF^{2}_{X}}
\newcommand{\cstib}{F^{2}_{\pi}F^{2}_{X}}
\newcommand{\ord}{\mathcal{O}}
\newcommand{\unme}{\frac{1}{2}}
\newcommand{\qu} {\mathbf{Q}}
\newcommand{\eps}{\varepsilon^{\mu\nu\rho\sigma}}

\newcommand{\runte}{\sqrt{\frac{1}{3}}}
\newcommand{\cstit}{F^{2}_{\pi}+3F^{2}_{X}}

\begin{document}

\addtolength{\baselineskip}{.8mm}

{\thispagestyle{empty}
\noindent \hspace{1cm}  \hfill Revised version \hspace{1cm}\\
\mbox{}                 \hfill June 2003 \hspace{1cm}\\

\begin{center}
\vspace*{1.0cm}
{\large\bf Reviewing the problem of the $U(1)$ axial symmetry}\\
{\large\bf and the chiral transition in QCD} \\
\vspace*{1.0cm}
{\large M. Marchi, E. Meggiolaro}\\
\vspace*{0.5cm}{\normalsize
{Dipartimento di Fisica, \\
Universit\`a di Pisa, \\
Via Buonarroti 2, \\
I--56127 Pisa, Italy.}}\\
\vspace*{2cm}{\large \bf Abstract}
\end{center}

\noindent
We discuss the role of the $U(1)$ axial symmetry for the phase structure of
QCD at finite temperature. We expect that, above a certain critical
temperature, also the $U(1)$ axial symmetry will be (effectively) restored.
We will try to see if this transition has (or has not) anything to do
with the usual chiral transition: various possible scenarios are discussed.
In particular, supported by recent lattice results, we analyse a scenario in
which a $U(1)$--breaking condensate survives across the chiral transition.
This scenario can be consistently reproduced using an effective Lagrangian
model. The effects of the $U(1)$ chiral condensate on the slope of the
topological susceptibility in the full theory with quarks are studied:
we find that this quantity (in the chiral limit of zero quark masses) acts as
an order parameter for the $U(1)$ axial symmetry above the chiral transition.
Further information on the new $U(1)$ chiral order parameter is derived
from the study (at zero temperature) of the radiative decays of the ``light''
pseudoscalar mesons in two photons: a comparison of our results with the
experimental data is performed.

\vspace{0.5cm}
\noindent
(PACS codes: 12.38.Aw, 12.39.Fe, 11.15.Pg, 11.30.Rd)
}
\vfill\eject

\newsection{Introduction}

\noindent
It is generally believed that a phase transition which occurs in QCD at a
finite temperature is the restoration of the spontaneously broken
$SU(L) \otimes SU(L)$ chiral symmetry in association with $L$ massless quarks.
At zero temperature the chiral symmetry is broken spontaneously by the
condensation of $q\bar{q}$ pairs and the $L^2-1$ $J^P=0^-$ mesons
are just the Goldstone bosons associated with this breaking
\cite{chiral-symmetry}.
At high temperatures the thermal energy breaks up the $q\bar{q}$ condensate,
leading to the restoration of chiral symmetry. We expect that this property
not only holds for massless quarks but also continues for a small mass region.
The order parameter for the chiral symmetry breaking is apparently
$\langle \bar{q}q \rangle \equiv \sum_{i=1}^L \langle \bar{q}_i q_i \rangle$:
the chiral symmetry breaking corresponds to the non--vanishing of
$\langle \bar{q}q \rangle$ in the chiral limit $\sup(m_i) \to 0$.
From lattice determinations of the chiral order parameter
$\langle \bar{q}q \rangle$ one knows that the $SU(L) \otimes SU(L)$ chiral
phase transition temperature $T_{ch}$, defined as the temperature at which
the {\it chiral condensate} $\langle \bar{q}q \rangle$ goes to zero (in the
chiral limit $\sup(m_i) \to 0$), is nearly equal to the deconfining temperature
$T_c$ (see, e.g., Ref. \cite{Blum-et-al.95}).
But this is not the whole story: QCD possesses not only an
approximate $SU(L) \otimes SU(L)$ chiral symmetry, for $L$ light quark
flavours, but also a $U(1)$ axial symmetry (at least at the classical level)
\cite{Weinberg75,tHooft76}.
The role of the $U(1)$ symmetry for the finite temperature phase
structure has been so far not well studied and it is still an open question
of hadronic physics whether the fate of the $U(1)$ chiral symmetry of QCD has
or has not something to do with the fate of the $SU(L) \otimes SU(L)$ chiral
symmetry.

In the ``Witten--Veneziano mechanism'' \cite{Witten79a,Veneziano79}
for the resolution of the $U(1)$ problem, a fundamental role is played by
the so--called ``topological susceptibility'' in a QCD without
quarks, i.e., in a pure Yang--Mills (YM) theory, in the large--$N_c$ limit
($N_c$ being the number of colours):
\be
A = \displaystyle\lim_{k \to 0}
\displaystyle\lim_{N_c \to \infty}
\left\{ -i \displaystyle\int d^4 x~ e^{ikx} \langle T Q(x) Q(0) \rangle
\right\} ,
\label{eqn1}
\ee
where $Q(x) = {g^2 \over 64\pi^2}\varepsilon^{\mu\nu\rho\sigma} F^a_{\mu\nu}
F^a_{\rho\sigma}$ is the so--called ``topological charge density''.
This quantity enters into the expression for the mass of the $\eta'$.
Therefore, in order to study the role of the $U(1)$ axial symmetry for the
full theory at non--zero temperatures, one should consider the YM topological
susceptibility $A(T)$ at a given temperature $T$, formally defined as in
Eq. (\ref{eqn1}), where now $\langle \ldots \rangle$ stands for the
expectation value in the full theory at the temperature $T$ \cite{EM1998}.

The problem of studying the behaviour of $A(T)$ as a function of the
temperature $T$ was first addressed, in lattice QCD,
in Refs. \cite{Teper86,EM1992a,EM1995b}.
Recent lattice results \cite{Alles-et-al.97} (obtained for the $SU(3)$
pure--gauge theory) show that the YM topological susceptibility $A(T)$
is approximately constant up to the critical temperature $T_c \simeq T_{ch}$,
it has a sharp decrease above the transition, but it remains different
from zero up to $\sim 1.2~T_c$. We recall that, in the Witten--Veneziano
mechanism \cite{Witten79a,Veneziano79}, a (no matter how small!) value
different from zero for $A$ is related to the breaking of the $U(1)$ axial
symmetry, since it implies the existence of a {\it would--be} Goldstone
particle with the same quantum numbers of the $\eta'$ (see also Ref.
\cite{Veneziano80}).

Another way to address the same question is to look at the behaviour at
non--zero temperatures of the susceptibilities related to the
propagators for the following meson channels \cite{Shuryak94}
(we consider for simplicity the case of $L=2$ light flavours):
the isoscalar $I=0$ scalar channel $\sigma$ (also known as $f_0$ in the
modern language of hadron spectroscopy), interpolated by the operator
$O_\sigma = \bar{q} q$;
the isovector $I=1$ scalar channel $\delta$
(also known as $a_0$), interpolated by the operator
$\vec{O}_\delta = \bar{q} {\vec{\tau} \over 2} q$;
the isovector $I=1$ pseudoscalar channel $\pi$, interpolated by the operator
$\vec{O}_\pi = i\bar{q} \gamma_5 {\vec{\tau} \over 2} q$;
the isoscalar $I=0$ pseudoscalar channel $\eta'$, interpolated by the operator
$O_{\eta'} = i\bar{q} \gamma_5 q$.
Under $SU(2)_A$ transformations, $\sigma$ is mixed with $\pi$: thus the
restoration of this symmetry at $T_{ch}$ requires identical correlators
for these two channels. Another $SU(2)$ chiral multiplet is $(\delta,\eta')$.
On the contrary, under the $U(1)_A$ transformations, $\pi$ is mixed
with $\delta$: so, an ``effective restoration'' of the $U(1)$ axial
symmetry should imply that these two channels become degenerate, with
identical correlators. Another $U(1)$ chiral multiplet is $(\sigma,\eta')$.
(Clearly, if both chiral symmetries are restored, then all $\pi$, $\eta'$,
$\sigma$ and $\delta$ correlators should become the same.)
In practice, one can construct, for each meson channel $f$, the
corresponding chiral susceptibility
\be
\chi_f = \displaystyle\int d^4x~ \langle O_f(x) O_f^\dagger(0) \rangle ,
\label{eqn2}
\ee
and then define two order parameters:
$\chi_{SU(2) \otimes SU(2)} \equiv \chi_\sigma - \chi_\pi$, and
$\chi_{U(1)} \equiv \chi_\delta - \chi_\pi$.
If an order parameter is non--zero in the chiral limit, then the
corresponding symmetry is broken.
Present lattice data for these quantities seem to indicate that the $U(1)$
order parameter survives across $T_{ch}$, up to $\sim 1.2~T_{ch}$,
where the $\delta$--$\pi$ splitting is small but still
different from zero \cite{Bernard-et-al.97,Karsch00,Vranas00}.
In terms of the left--handed and right--handed quark
fields,\footnote{Throughout this paper we use the following notations:
$q_{L,R} \equiv {1 \over 2} (1 \pm \gamma_5) q$, with $\gamma_5 \equiv
-i\gamma^0\gamma^1\gamma^2\gamma^3$.}
one has the following expression for the difference between the correlators
for the $\delta^+$ and $\pi^+$ channels:
\ba
\lefteqn{
{\cal D}_{U(1)}(x) \equiv \langle O_{\delta^+}(x) O_{\delta^+}^\dagger(0)
\rangle - \langle O_{\pi^+}(x) O_{\pi^+}^\dagger(0) \rangle } \nonumber \\
& & = 2 \left[ \langle \bar{u}_R d_L(x) \cdot \bar{d}_R u_L(0) \rangle
+ \langle \bar{u}_L d_R(x) \cdot \bar{d}_L u_R(0) \rangle \right] .
\label{eqn3}
\ea
(The integral of this quantity, $\int d^4x~ {\cal D}_{U(1)}(x)$, is just equal
to the $U(1)$ chiral susceptibility $\chi_{U(1)} = \chi_\delta - \chi_\pi$.)
What happens below and above $T_{ch}$?
Below $T_{ch}$, in the chiral limit $\sup(m_i) \to 0$, the left--handed
and right--handed components of a given light quark flavour ({\it up} or
{\it down}, in our case with $L=2$) can be connected through the $q\bar{q}$
chiral condensate, giving rise to a non--zero contribution to the
quantity ${\cal D}_{U(1)}(x)$ in Eq. (\ref{eqn3}) (i.e., to the quantity
$\chi_{U(1)}$). But above $T_{ch}$ the $q\bar{q}$ chiral condensate is zero:
so, how can the quantity
${\cal D}_{U(1)}(x)$ (i.e., the quantity $\chi_{U(1)}$) be different from zero
also above $T_{ch}$, as indicated by present lattice data?
The only possibility in order to solve this puzzle seems to be that of
requiring the existence of a genuine four--fermion local condensate,
which is an order parameter for the $U(1)$ axial symmetry and which
remains different from zero also above $T_{ch}$.
This new condensate will be discussed in Section 3. The rest of the paper
will be essentially devoted to the analysis of some interesting
phenomenological consequences deriving from this hypothesis.

The paper is organized as follows.
In Section 2 we discuss the role of the $U(1)$ axial
symmetry for the phase structure of QCD at finite temperature.
One expects that, above a certain critical temperature, also the $U(1)$ axial
symmetry will be (effectively) restored. We will try to see if this transition
has (or has not) anything to do with the usual chiral transition: various
possible scenarios are discussed.
In particular, supported by the above--mentioned lattice results, in Sections
3 and 4 we analyse a scenario in which a new $U(1)$--breaking condensate
survives across the chiral transition and it is still present above $T_{ch}$.
This scenario can be consistently reproduced using an effective Lagrangian
model, which also includes the new $U(1)$ chiral condensate.
This theoretical model was originally proposed in Refs.
\cite{EM1994a,EM1994b,EM1994c} and is summarized in Sections 3 and 4 for the
convenience of the reader.
(See also Refs. \cite{EM2002a,EM2002b} for a recent review on these problems.)
In Section 5 (which, together with Section 6, contains the main original
results of this paper) we analyse the consequences of our theoretical model
on the slope of the topological susceptibility $\chi'$, in the {\it full}
theory with quarks, showing how this quantity is modified by the presence
of a new $U(1)$ chiral order parameter: we will find that $\chi'$
(in the chiral limit $\sup(m_i) \to 0$) acts as an order parameter
for the $U(1)$ axial symmetry above $T_{ch}$.
Further information on the new $U(1)$ chiral order parameter is derived in
Section 6 from the study (at zero temperature) of the radiative decays of the
``light'' pseudoscalar mesons in two photons: a comparison of our results
with the experimental data is performed.
Finally, the conclusions and an outlook are given in Section 7.

\newsection{The phase transitions of QCD}

\noindent
One expects that, above a certain critical temperature, also the $U(1)$ axial
symmetry will be (effectively) restored.\footnote{Of course, the $U(1)$ axial
symmetry is explicitly broken by the anomaly, which never disappears: the
precise meaning of this ``effective restoration'' will be explained below.}
We will try to see if this transition has (or has not) anything to do with
the usual chiral transition. Let us define the following temperatures:
\begin{itemize}
\item{} $T_{ch}$: the temperature at which the chiral condensate
$\langle \bar{q} q \rangle$ goes to zero. The chiral symmetry
$SU(L) \otimes SU(L)$ is spontaneously broken below $T_{ch}$ and
it is restored above $T_{ch}$.
\item{} $T_\chi$: the temperature at which the pure--gauge topological
susceptibility $A$ drops to zero. Present lattice results
indicate that $T_\chi \ge T_{ch}$ \cite{Alles-et-al.97}.
\item{} $T_{U(1)}$: the temperature at which the $U(1)$ axial symmetry
is (effectively) restored, meaning that, for $T>T_{U(1)}$, there are no
$U(1)$--breaking condensates.
If $\langle \bar{q} q \rangle \ne 0$ also the $U(1)$ axial symmetry is
broken, i.e., the chiral condensate is an order parameter also for
the $U(1)$ axial symmetry. Therefore we must have: $T_{U(1)} \ge T_{ch}$.
Moreover, the Witten--Veneziano mechanism implies that $T_{U(1)} \ge T_\chi$,
since, after all, the pure--YM topological susceptibility $A$ is a
$U(1)$--breaking condensate.
\end{itemize}
The following scenario, that we will call ``\underline{SCENARIO 1}'',
in which $T_\chi < T_{ch}$, is, therefore, immediately ruled out.
In this case, in the range of temperatures between $T_\chi$ and $T_{ch}$
the anomaly effects are absent, but the $U(1)$ axial symmetry is still
broken by the chiral condensate. In other words, in this range of temperatures
the $U(1)$ axial symmetry is spontaneously broken ({\it \`a la} Goldstone)
and the $\eta'$ is the corresponding Goldstone boson, i.e., it is massless in
the chiral limit $\sup(m_i) \to 0$, or, at least, as light as the pion $\pi$,
when including the quark masses.
This scenario was first discussed (and indeed really supported!) in
Ref. \cite{Pisarski-Wilczek84}.

Therefore, we are left essentially with the two following scenarios.\\
\underline{SCENARIO 2}: $T_{ch} \le T_{U(1)}$, with
$T_{ch} \sim T_\chi \sim T_{U(1)}$.
If $T_{ch} = T_\chi = T_{U(1)}$, then, in the case of $L=2$ light flavours,
the restored symmetry across the transition is $U(1)_A \otimes SU(2)_L \otimes
SU(2)_R \sim O(2) \otimes O(4)$, which may yield a first--order phase
transition (see, for example, Ref. \cite{Kharzeev-et-al.98}).\\
\underline{SCENARIO 3}: $T_{ch} \ll T_{U(1)}$, that is, the complete
$U(L)_L \otimes U(L)_R$ chiral symmetry is restored only well inside the
quark--gluon plasma domain.
In the case of $L=2$ light flavours, we then have at $T=T_{ch}$ the
restoration of $SU(2)_L \otimes SU(2)_R \sim O(4)$.
Therefore, we can have a second--order phase transition with the
$O(4)$ critical exponents. $L=2$ QCD at $T \simeq T_{ch}$ and the $O(4)$
spin system should belong to the same universality class.
An effective Lagrangian describing the softest modes is essentially
the Gell-Mann--Levy linear sigma model, the same as for the $O(4)$
spin systems (see Ref. \cite{Pisarski-Wilczek84}).
If this scenario is true, one should find the $O(4)$ critical indices
for the $q\bar{q}$ chiral condensate and the specific heat:
$\langle \bar{q} q \rangle \sim |(T - T_{ch})/T_{ch}|^{0.38 \pm 0.01}$,
and $C(T) \sim |(T - T_{ch})/T_{ch}|^{0.19 \pm 0.06}$.
Present lattice data partially support these results.

\newsection{The $U(1)$ chiral order parameter}

\noindent
We make the assumption (discussed in the previous sections) that the
breaking/restoration of the $U(1)$ chiral symmetry is completely independent
of the $SU(L) \otimes SU(L)$ symmetry.
The usual chiral order parameter $\langle \bar{q} q \rangle$
is an order parameter both for $SU(L) \otimes SU(L)$ and for $U(1)_A$:
when it is different from zero, $SU(L) \otimes SU(L)$ is broken down to
$SU(L)_V$ and also $U(1)_A$ is broken.
Thus we need another quantity which could be an order parameter only for
the $U(1)$ chiral symmetry \cite{EM1994a,EM1994b,EM1994c,EM1995a}.
The most simple quantity of this kind was found by 'tHooft
in Ref. \cite{tHooft76}.
For a theory with $L$ light quark flavours, it is a $2L$--fermion interaction
that has the chiral transformation properties of:
\be
{\cal L}_{eff} \sim \displaystyle{{\det_{st}}(\bar{q}_{sR}q_{tL})
+ {\det_{st}}(\bar{q}_{sL}q_{tR}) },
\label{eqn5}
\ee
where $s,t = 1, \ldots ,L$ are flavour indices, but the colour indices are
arranged in a more general way (see Refs. \cite{EM1994c,EM1995a}).
It is easy to verify that ${\cal L}_{eff}$ is invariant under
$SU(L) \otimes SU(L) \otimes U(1)_V$, while it is not invariant under $U(1)_A$.
To obtain an order parameter for the $U(1)$ chiral symmetry, one can
simply take the vacuum expectation value of ${\cal L}_{eff}$:
$C_{U(1)} = \langle {\cal L}_{eff} \rangle$.
The arbitrarity in the arrangement of the colour indices can be removed if we
require that the new $U(1)$ chiral condensate is ``independent'' of the
usual chiral condensate $\langle \bar{q} q \rangle$, as explained in
Refs. \cite{EM1994c,EM1995a}. In other words, the condensate $C_{U(1)}$
is chosen to be a {\it genuine} $2L$--fermion condensate, with a zero
``disconnected part'', the latter being the contribution proportional
to $\langle \bar{q} q \rangle^L$,
corresponding to retaining the vacuum intermediate state in all the channels
and neglecting the contributions of all the other states.
As a remark, we observe that the condensate $C_{U(1)}$ so defined
turns out to be of order ${\cal O}(g^{2L - 2} N_c^L) = {\cal O}(N_c)$
in the large--$N_c$ expansion, exactly as the chiral condensate
$\langle \bar{q} q \rangle$.

The existence of a new $U(1)$ chiral order parameter has of course interesting
physical consequences, which can be revealed by analysing some relevant QCD
Ward Identities (WI's) (see Ref. \cite{EM1994b} and also Ref. \cite{EM2002a}).
In the case of the $SU(L) \otimes SU(L)$ chiral symmetry, one
immediately derives the following WI:
\be
\int d^4 x \langle T\partial^\mu A^a_\mu (x)
i\bar{q} \gamma_5 T^b q(0) \rangle
= i \delta_{ab} {1 \over L} \langle \bar{q} q \rangle ,
\label{eqn6}
\ee
where $A^a_\mu = \bar{q}\gamma_\mu \gamma_5 T^a q$ are the $SU(L)$ axial
currents. If $\langle \bar{q} q \rangle \ne 0$ (in the chiral limit
$\sup(m_i) \to 0$), the anomaly--free WI (\ref{eqn6}) implies the existence
of $L^2-1$ non--singlet Goldstone bosons, interpolated by the hermitian fields
$O_b = i \bar{q} \gamma_5 T^b q$.
Similarly, in the case of the $U(1)$ axial symmetry, one finds that:
\be
\int d^4x~ \langle T\partial^\mu J_{5, \mu}(x) i\bar{q} \gamma_5 q(0)
\rangle = 2i \langle \bar{q} q \rangle ,
\label{eqn7}
\ee
where $J_{5, \mu}= {\bar{q} \gamma_\mu \gamma_5 q}$ is the $U(1)$ axial
current. But this is not the whole story! One also derives the following WI:
\be
\int d^4x~ \langle T\partial^\mu J_{5, \mu}(x) O_P(0) \rangle =
2Li \langle {\cal L}_{eff}(0) \rangle ,
\label{EQN10.5}
\ee
where ${\cal L}_{eff}$ is the $2L$--fermion operator defined by
Eq. (\ref{eqn5}), while the hermitian field $O_P$ is defined as:
$O_P \sim i[ {\det} (\bar{q}_{sR}q_{tL}) - {\det} (\bar{q}_{sL}q_{tR}) ]$.
If the $U(1)$--breaking condensate survives across the chiral transition at
$T_{ch}$, i.e., $C_{U(1)} = \langle {\cal L}_{eff}(0) \rangle \ne 0$ for
$T > T_{ch}$ (while $\langle \bar{q} q \rangle = 0$ for $T > T_{ch}$), then
this WI implies the existence of a ({\it would--be}) Goldstone boson (in the
large--$N_c$ limit) coming from this breaking and interpolated by the hermitian
field $O_P$. Therefore, the $U(1)_A$ ({\it would--be}) Goldstone boson (i.e.,
the $\eta'$) is an ``exotic'' $2L$--fermion state for $T > T_{ch}$.

\newsection{The new chiral effective Lagrangian}

\noindent
The proposed scenario, in which the $U(1)$ axial symmetry is (effectively)
restored at a temperature $T_{U(1)}$ greater than $T_{ch}$, can be consistently
reproduced using an effective--Lagrangian model. This analysis was originally
performed in Refs. \cite{EM1994a,EM1994b,EM1994c} and is summarized here for
the convenience of the reader.

It is well known that the low--energy dynamics of the pseudoscalar mesons,
including the effects due to the anomaly and the $q\bar{q}$ chiral condensate,
and expanding to the first order in the light quark masses, can be described,
in the large--$N_c$ limit, by an effective Lagrangian
\cite{DiVecchia-Veneziano80,Witten80,Rosenzweig-et-al.80,Nath-Arnowitt81,Ohta80}
written in terms of the mesonic field $U_{ij} \sim \bar{q}_{jR} q_{iL}$
(up to a multiplicative constant) and the topological charge density $Q$.
We make the assumption that there is a $U(1)$--breaking condensate which
stays different from zero across $T_{ch}$, up to $T_{U(1)} > T_{ch}$:
the form of this condensate has been discussed in the previous section.
We must now define a field variable $X$,
associated with this new condensate, to be inserted in the chiral Lagrangian.
The translation from the fundamental quark fields to the
effective--Lagrangian meson fields is done as follows. The operators
$i \bar{q} \gamma_5 q$ and $\bar{q} q$ entering in the WI (\ref{eqn7})
are essentially equal to (up to a multiplicative constant)
$i(\Tr U - \Tr U^\dagger)$ and $\Tr U + \Tr U^\dagger$ respectively.
Similarly, the operators
${\cal L}_{eff} \sim {\det}(\bar{q}_{sR}q_{tL})
+ {\det}(\bar{q}_{sL}q_{tR})$ and
$O_P \sim i[ {\det}(\bar{q}_{sR}q_{tL})
- {\det}(\bar{q}_{sL}q_{tR}) ]$
entering in the WI (\ref{EQN10.5}) can be put equal to (up to a multiplicative
constant) $X + X^\dagger$ and $i(X - X^\dagger)$ respectively, where
$X \sim {\det} \left( \bar{q}_{sR} q_{tL} \right)$
is the new field variable (up to a multiplicative constant),
related to the new $U(1)$ chiral condensate, which must be inserted
in the chiral effective Lagrangian.
It was shown in Refs. \cite{EM1994a,EM1994b,EM1994c} that the most simple
effective Lagrangian, constructed with the fields $U$, $X$ and $Q$, is:
\ba
\lefteqn{
{\cal L}(U,U^\dagger ,X,X^\dagger ,Q) =
{1 \over 2}\Tr(\partial_\mu U\partial^\mu U^\dagger )
+ {1 \over 2}\partial_\mu X\partial^\mu X^\dagger } \nonumber \\
& & -V(U,U^\dagger ,X,X^\dagger)
+{1 \over 2}iQ(x)\omega_1 \Tr(\ln U - \ln U^\dagger) \nonumber \\
& & +{1 \over 2}iQ(x)(1-\omega_1)(\ln X-\ln X^\dagger)+{1 \over 2A}Q^2(x),
\label{eqn9}
\ea
where the potential term $V(U,U^{\dagger},X,X^{\dagger})$ has the form:
\ba
\lefteqn{
V(U,U^\dagger ,X,X^\dagger )={1 \over 4}\lambda_{\pi}^2 \Tr[(U^\dagger U
-\rho_\pi \cdot {\bf I})^2] +
{1 \over 4}\lambda_X^2 (X^\dagger X-\rho_X )^2 } \nonumber \\
& & -{B_m \over 2\sqrt{2}}\Tr(MU+M^\dagger U^\dagger)
-{c_1 \over 2\sqrt{2}}[\det(U)X^\dagger + \det(U^\dagger )X].
\label{eqn10}
\ea
{\bf I} is the identity matrix.
$M$ represents the quark mass matrix, $M={\rm diag}(m_1,\ldots,m_L)$,
which enters in the QCD Lagrangian as $\delta {\cal L}^{(mass)}_{QCD} =
-\bar{q}_R M q_L -\bar{q}_L M^\dagger q_R$,
and $A$ is the topological susceptibility in the pure--YM theory.
All the parameters appearing in the Lagrangian must be considered as
functions of the physical temperature $T$. In particular, the parameters
$\rho_{\pi}$ and $\rho_X$ are responsible for the behaviour of the theory
respectively across the $SU(L) \otimes SU(L)$ and the $U(1)$ chiral phase
transitions, as follows:
\ba
\rho_\pi(T<T_{ch}) &\equiv& {1 \over 2} F_\pi^2 > 0, ~~~
\rho_\pi(T>T_{ch}) < 0; \nonumber \\
\rho_X(T<T_{U(1)}) &\equiv& {1 \over 2} F_X^2 > 0, ~~~
\rho_X(T>T_{U(1)}) < 0.
\label{table}
\ea
The parameter $F_\pi$ is the well--known pion decay constant, while the
parameter $F_X$ is related to the new $U(1)$ axial condensate and will be
discussed at length in the rest of the paper.
For $T<T_{ch}$, $\rho_\pi > 0$ and therefore, by virtue of the form
(\ref{eqn10}) of the potential,\footnote{As it has been stressed in Ref.
\cite{EM1994a} (see also Ref. \cite{DiVecchia-Veneziano80}), the {\it linear
$\sigma$--type} model (\ref{eqn9})--(\ref{eqn10}) contains {\it redundant}
scalar fields (we are only interested in the pseudoscalar Goldstone, or
{\it would--be} Goldstone, bosons), which can be eliminated by taking the
limits $\lambda_\pi^2,\lambda_X^2 \to +\infty$ in Eq. (\ref{eqn10}):
so an expansion is performed not only in powers of the light quark masses
$m_i$, but also in powers of $1/\lambda_\pi^2$ and $1/\lambda_X^2$.}
one finds that $\langle U \rangle \ne 0$,
or, in other words $\langle \bar{q} q \rangle \ne 0$ (being
$U_{ij} \sim \bar{q}_{jR} q_{iL}$, up to a multiplicative constant):
i.e., the $SU(L) \otimes SU(L)$ chiral symmetry is broken.
Instead, for $T>T_{ch}$, $\rho_\pi < 0$ and then, always from
Eq. (\ref{eqn10}), one has that $\langle U \rangle = 0$, i.e.,
$\langle \bar{q} q \rangle = 0$.
The $U(1)$ chiral condensate stays different from zero also in the region of
temperatures $T_{ch} < T < T_{U(1)}$, where, on the contrary, the
$SU(L) \otimes SU(L)$ chiral symmetry is restored.
In fact, for $T<T_{U(1)}$, $\rho_X > 0$ and therefore, from Eq. (\ref{eqn10}),
one finds that $\langle X \rangle \ne 0$, or, in other words, $C_{U(1)} =
\langle {\cal L}_{eff} \rangle \ne 0$ (being
$X \sim {\det} \left( \bar{q}_{sR} q_{tL} \right)$, up to a multiplicative
constant). The $U(1)$ chiral symmetry is (effectively) restored above
$T_{U(1)}$, where $\rho_X < 0$ and then, from Eq. (\ref{eqn10}),
$\langle X \rangle = 0$, i.e., $C_{U(1)} = \langle {\cal L}_{eff} \rangle = 0$.
According to what we have said in the Introduction and in Section 2,
we also assume that the topological susceptibility $A(T)$ of the pure--YM
theory drops to zero at a temperature $T_{\chi}$ greater than $T_{ch}$
(but smaller than, or equal to, $T_{U(1)}$).

One can study the mass spectrum of the theory for $T < T_{ch}$ and
$T_{ch} < T < T_{U(1)}$. First of all, let us see what happens for
$T<T_{ch}$, where both the $q\bar{q}$ chiral condensate and the $U(1)$ chiral
condensate are present. Integrating out the field variable $Q$ and taking only
the quadratic part of the Lagrangian, one finds that, in the chiral limit
$\sup(m_i) \to 0$, there are $L^2-1$ zero--mass states, which represent the
$L^2-1$ Goldstone bosons coming from the breaking of the $SU(L) \otimes SU(L)$
chiral symmetry down to $SU(L)_V$. Then there are two singlet eigenstates
with non--zero masses:
\ba
\eta' &=& {1 \over \sqrt{F_\pi^2 + LF_X^2}}(\sqrt{L}F_X S_X + F_\pi S_\pi),
\nonumber \\
\eta_X &=& {1 \over \sqrt{F_\pi^2 + LF_X^2}}(-F_\pi S_X + \sqrt{L}F_X S_\pi),
\label{eqn11}
\ea
where $S_\pi$ is the usual $SU(L)$--singlet meson field associated with $U$,
while $S_X$ is the meson field associated with $X$ (see Refs.
\cite{EM1994a,EM1994b,EM1994c} and Eqs. (\ref{u,x}) below).
The field $\eta'$ has a ``light'' mass, in the sense of the
$N_c \to \infty$ limit, being
\be
m^2_{\eta'} = {2LA \over F_\pi^2 + LF_X^2} = {\cal O}({1 \over N_c}).
\label{eqn12}
\ee
This mass is intimately related to the anomaly and they both vanish in the
$N_c \to \infty$ limit. On the contrary, the field $\eta_X$ has a sort of
``heavy hadronic'' mass of order ${\cal O}(N_c^0)$ in the large--$N_c$ limit.
We immediately see that, if we put $F_X=0$ in the above--written
formulae (i.e., if we neglect the new $U(1)$ chiral condensate), then
$\eta' = S_\pi$ and $m^2_{\eta'}$ reduces to ${2LA \over F^2_\pi}$, which is
the ``usual'' $\eta'$ mass in the chiral limit \cite{Witten79a,Veneziano79}.
Yet, in the general case $F_X \ne 0$, the two states which diagonalize the
squared mass matrix are linear combinations of the ``quark--antiquark''
singlet field $S_\pi$ and of the ``exotic'' field $S_X$.
Both the $\eta'$ and the $\eta_X$ have the same quantum numbers (spin,
parity and so on), but they have a different quark content: one is mostly
$\sim i(\bar{q}_{L}q_{R}-\bar{q}_{R}q_{L})$, while the other is mostly
$\sim i[ {\det}(\bar{q}_{sL}q_{tR}) - {\det}(\bar{q}_{sR}q_{tL}) ]$.
What happens when approaching the chiral transition temperature $T_{ch}$?
We know that $F_\pi(T) \to 0$ when $T \to T_{ch}$. From Eq. (\ref{eqn12})
we see that $m^2_{\eta'}(T_{ch}) = {2A \over F_X^2}$
and, from the first Eq. (\ref{eqn11}), $\eta'(T_{ch}) = S_X$.
We have continuity in the mass spectrum of the theory through the chiral
phase transition at $T=T_{ch}$.
In fact, if we study the mass spectrum of the theory in the region of
temperatures $T_{ch} < T < T_{U(1)}$ (where the $SU(L) \otimes SU(L)$ chiral
symmetry is restored, while the $U(1)$ chiral condensate is still present),
one finds that there is a singlet meson field $S_X$ (associated with the
field $X$ in the chiral Lagrangian) with a squared mass given by (in the
chiral limit): $m^2_{S_X} = {2A \over F_X^2}$.
This is nothing but the {\it would--be} Goldstone particle
coming from the breaking of the $U(1)$ chiral symmetry, i.e., the $\eta'$,
which, for $T>T_{ch}$, is a sort of ``exotic'' matter field of the form
$\sim i[ {\det}(\bar{q}_{sL}q_{tR})
- {\det}(\bar{q}_{sR}q_{tL}) ]$.
Its existence could be proved perhaps in the near
future by heavy--ion experiments.

\newsection{A relation between $\chi'$ and the new $U(1)$ chiral condensate}

\noindent
In this section and in the following one we want to describe some methods which
provide us with some information about the parameter $F_X$. This quantity
is a $U(1)$--breaking parameter: indeed, from Eq. (\ref{table}),
$\rho_X = {1 \over 2} F_X^2 > 0$ for $T<T_{U(1)}$, and therefore, from Eq.
(\ref{eqn10}), $\langle X \rangle = F_X/\sqrt{2} \ne 0$. Remembering that
$X \sim {\det} \left( \bar{q}_{sR} q_{tL} \right)$, up to a multiplicative
constant, we find that $F_X$ is proportional to the new $2L$--fermion
condensate $C_{U(1)} = \langle {\cal L}_{eff} \rangle$ introduced above.\\
In the same way, the pion decay constant $F_{\pi}$, which controls the breaking
of the $\sgru$ symmetry, is related to the $q\bar{q}$ chiral condensate
by a simple and well--known proportionality relation (see Ref. \cite{EM1994a}
and references therein):
$\langle \bar{q}_i q_i \rangle_{T<T_{ch}} \simeq -{1 \over 2}B_m F_\pi$.
Considering, for simplicity, the case of $L$ light quarks with the same mass
$m$, one immediately derives from this equation the so--called
\emph{Gell-Mann--Oakes--Renner relation} \cite{GOR68}:
\be
\label{GOR}
m_{NS}^{2}F_{\pi}^{2}\simeq-\frac{2m}{L}\langle\bar{q}q\rangle_{T<T_{ch}},
\ee
where, as usual,
$\langle \bar{q}q \rangle \equiv \sum_{i=1}^L \langle \bar{q}_i q_i \rangle$,
and, moreover, $m^2_{NS} = m B_m/F_\pi$, $m_{NS}$ being the mass of the
non--singlet pseudoscalar mesons.
Eq. (\ref{GOR}) relates, on the left--hand side, the pion decay constant
$F_{\pi}$ and the mass $m_{NS}$ of the non--singlet mesons with, on the
right--hand side, the chiral condensate $\langle\bar{q}q\rangle$ and the
quark mass $m$.\\
It is not possible to find, in a simple way, the analogous relation between
$F_X$ and the new condensate $C_{U(1)} = \langle {\cal L}_{eff} \rangle$,
since the QCD Lagrangian does not contain any term proportional to the
$2L$--fermion operator ${\cal L}_{eff}$.

Alternatively, the quantity $F_{X}$ can be written in terms of a certain
two--point Green function of the topological charge--density operator
$Q(x)$ in the {\it full} theory with quarks.\\
If we want to derive the two--point function of $Q(x)$, we need to consider the
effective Lagrangian in the form (\ref{eqn9}), where the field variable $Q(x)$
has not yet been integrated. Therefore:
\be
\chi(k)\equiv-i\int d^{4}x\;e^{ikx}\langle{TQ(x)Q(0)}\rangle=
({\cal K}^{-1}(k))_{11},
\label{chik}
\ee
where ${\cal K}^{-1}(k)$ is the inverse of the matrix ${\cal K}(k)$ associated
with the quadratic part of the Lagrangian (\ref{eqn9}) in the momentum space,
for the ensemble of fields $(Q(x),\ldots )$.\\
In particular, for $T<T_{ch}$, one has to consider the following quadratic
Lagrangian, in the chiral limit $\sup(m_i) \to 0$:
\ba
\label{l2ch}
\lefteqn{\La_{2}=
\unme\sum_{a=1}^{L^2-1}\partial_{\mu}\pi_{a}\partial^{\mu}\pi_{a}+
\unme \partial_{\mu}S_{\pi}\partial^{\mu}S_{\pi}+
\unme \partial_{\mu}S_{X}\partial^{\mu}S_{X} }&&\nonumber\\
&&-\unme c\Bigg(\frac{\sqrt{2L}}{F_{\pi}}S_{\pi}
-\frac{\sqrt{2}}{F_{X}}S_{X}\Bigg)^{2} +\frac{1}{2A}Q^{2} \nonumber \\
&& -\omega_{1}\frac{\sqrt{2L}}{F_{\pi}}S_{\pi}Q
-(1-\omega_{1})\frac{\sqrt2}{F_{X}}S_{X}Q,
\ea
where $c \equiv {c_1 \over \sqrt{2}} \left({F_X \over \sqrt{2}}\right)
\left({F_\pi \over \sqrt{2}}\right)^L$.
We have used for $U$ and $X$ the following exponential form, valid for
$T<T_{ch}$ \cite{EM1994a,EM1994b,EM1994c}:
\ba
U &=& \frac{F_\pi}{\sqrt2}\exp\left( {i\sqrt{2}\over F_\pi}\Phi\right),~~~
{\rm with:}~~ \Phi = \displaystyle\sum_{a=1}^{L^2-1}
\pi_{a}\tau_{a}+\frac{S_{\pi}}{\sqrt L}\cdot\I ;
\nonumber \\
X &=& \frac{F_X}{\sqrt2}\exp\left({i\sqrt{2}\over F_X} S_X\right),
\label{u,x}
\ea
where the matrices $\tau_a$ ($a=1,\ldots,L^2-1$) are the generators of the
algebra of $SU(L)$ in the fundamental representation, with normalization:
$\Tr(\tau_a) = 0\,,\;\;\Tr(\tau_a \tau_b) = \delta_{ab}.$\\
As already pointed out in Section 4, $S_\pi$ is the usual ``quark--antiquark''
$SU(L)$--singlet meson field associated with $U$, while $S_X$ is the ``exotic''
$2L$--fermion meson field associated with $X$.
The $\pi_a$ ($a=1,\ldots,L^2-1$) are the self--hermitian fields describing
the $L^2-1$ pions: they are massless in the chiral limit $\sup(m_i) \to 0$.\\
From the quadratic part of the Lagrangian (\ref{l2ch}) in the momentum space,
we derive the following matrix ${\cal K}(k)$ for the ensemble of fields
$(Q,S_X,S_{\pi})$ [the contribution of the pion fields $\pi_a$ is simply
diagonal, diag($k^2,\ldots,k^2$), and therefore can be trivially factorized
out]:
\be
\label{matrice1}
{\cal K}(k)=
\left( \begin{array}{ccc}
\frac{1}{A} & -\frac{\sqrt{2}(1-\omega_{1})}{F_{X}} &
-\frac{\omega_1 \sqrt{2L}}{F_{\pi}} \\
 & & \\
-\frac{\sqrt{2}(1-\omega_{1})}{F_{X}} & k^{2}-\frac{2c}{F^{2}_{X}} &
\frac{2c\sqrt{L}}{F_{\pi}F_{X}} \\
 & & \\
-\frac{\omega_1 \sqrt{2L}}{F_{\pi}} & \frac{2c\sqrt{L}}{F_{\pi}F_{X}} &
k^{2}-\frac{2Lc}{F^{2}_{\pi}}
\end{array} \right).
\ee
The calculation of the right--hand side of Eq. (\ref{chik}) can then be
performed explicitly, using Eq. (\ref{matrice1}), obtaining:
\ba
\label{chisotto}
\chi(k)=\frac{A\Big[k^{4}-\frac{2c(\csti)}{\cstib}k^{2}\Big]}
{\Big[k^{4}-\frac{2c(\csti)}{\cstib}k^{2}\Big]
-2A\Big[\frac{(1-\omega_{1})^{2}F_{\pi}^{2}+L\omega_{1}^{2}F_{X}^{2}}
{\cstib}\Big]k^{2}
+\frac{4LAc}{\cstib}}.
\ea
One immediately sees that $\chi \equiv \chi(0)=0$, i.e., the topological
susceptibility in the full theory with quarks vanishes in the chiral limit
$\sup(m_i) \to 0$, as expected.\\
But the most interesting result is found when considering the so--called
``slope'' of the topological susceptibility, defined as:
\ba
\label{chiprimo}
\chi'\equiv\frac{1}{8}{\partial \over \partial k_\mu}
{\partial \over \partial k^\mu} \chi(k)\Bigg|_{k=0}=
\frac{i}{8}\int d^{4}x \;x^{2} \langle TQ(x)Q(0)\rangle.
\ea
Moreover, whenever $\chi(k)$ is a function of $k^2$ (such as in the theory
at $T=0$, by virtue of the Lorentz invariance), one can also write:
\ba
\label{chiprimo-bis}
\chi'=\frac{d}{dk^{2}}\chi(k)\Bigg|_{k=0}.
\ea
In the theory at finite temperature $T \ne 0$ the Lorentz invariance is
broken down to the $O(3)$ invariance under spatial rotations only,
so that $\chi(k)$ is, in general, a function of $\vec{k}^2$ and $k^0$.
However, the propagator $\chi(k)$ obtained in Eq. (\ref{chisotto}) is a
function of $k^2 = (k^0)^2 - \vec{k}^2$ also in the case of non--zero
temperature. Therefore, we can explicitly calculate the quantity $\chi'$
in our effective model in the chiral limit $\sup(m_i) \to 0$ (and we shall
call ``$\chi'_{ch}$'' its value), using Eqs. (\ref{chisotto}) and
(\ref{chiprimo-bis}), obtaining:
\ba
\label{chi'sotto}
\chi'_{ch}=-\frac{1}{2L}(\csti)\equiv-\frac{1}{2L}F_{\eta'}^2,
\ea
where $F_{\eta'}\equiv\sqrt{\csti}$ is the decay constant of the $\eta'$ (at
the leading order in the $1/N_c$ expansion), modified by the presence of the
new $U(1)$ chiral order parameter \cite{EM1994c}.

For the benefit of the reader, we here briefly repeat the arguments developed
in Ref. \cite{EM1994c}, leading to the result $F_{\eta'}\equiv\sqrt{\csti}$:
this will also provide us with an alternative derivation of
Eq. (\ref{chi'sotto}). We shall consider the $T=0$ case for simplicity.\\
It turns out that $\eta'$ is just the meson state, with a squared mass of order
$1/N_c$, whose contribution to the full topological susceptibility $\chi$
exactly cancels out (in the chiral limit of massless quarks) the pure--gauge
part $A$ of $\chi$, so making $\chi = 0$: this is the so--called
Witten's mechanism. To see how this picture comes out in our theory, one first
determines the $U(1)$ axial current, starting from our effective Lagrangian.
This is easily done remembering how the fields $U$ and $X$ transform under a
$U(1)$ chiral transformation and one ends up with the following expression
\cite{EM1994c,EM2002a}:
\be
J_{5, \mu} = i[\Tr(U^\dagger \partial_\mu U - U \partial_\mu U^\dagger)
+ L(X^\dagger \partial_\mu X - X \partial_\mu X^\dagger)] .
\label{EQN5.15}
\ee
After having inserted here the expressions (\ref{u,x}) in place
of $U$ and $X$, the current $J_{5, \mu}$ takes the following form:
\be
J_{5, \mu} = -\sqrt{2L} F_{\eta'} \partial_\mu \eta' ,
\label{EQN5.17}
\ee
where the field $\eta'$ is defined by the first Eq. (\ref{eqn11}) and
the relative coupling between $J_{5, \mu}$ and $\eta'$, i.e., the
$SU(L)$--singlet ($\eta'$) decay constant defined as
$\langle 0|J_{5, \mu}(0)|\eta'(p)\rangle = i\sqrt{2L}\,p_\mu\,F_{\eta'}$,
is given by:
\be
F_{\eta'} = \sqrt{\csti} .
\label{EQN5.18}
\ee
Let us now recall the Witten's argument and write the two--point function (at
four--momentum $k$) of the topological charge density $Q(x)$ as a sum over
one--hadron poles, i.e., one--hadron intermediate states:
\be
\chi(k) = -i\int d^4x\, e^{ikx} \langle TQ(x)Q(0) \rangle = A_0(k)
+ \displaystyle\sum_{mesons}{|\langle 0|Q|n \rangle |^2 \over k^2 - m^2_n} ,
\label{EQN5.19}
\ee
where $A_0(k)$ is the pure Yang--Mills contribution from the glueball
intermediate states and it is the leading--order term in $1/N_c$ [being of
order $\ord(N_c^0)$].
In the chiral limit in which we have $L$ massless quarks, the full topological
susceptibility $\chi\equiv\chi(k=0)$ must vanish: so there must be a meson
state, with squared mass $m^2_n = {\cal O}(1/N_c)$ [since $A_0(0)=\ord(N_c^0)$,
while $|\langle 0|Q|n \rangle |^2 = {\cal O}(1/N_c)$], which exactly cancels
out $A_0(0)$. This is the meson that we usually call $\eta'$: the other meson
states in (\ref{EQN5.19}) have squared masses of order $\ord(N_c^0)$, so that
their contributions to the summation in (\ref{EQN5.19}) are suppressed by a
factor of $1/N_c$. Therefore we obtain that:
\be
{|\langle 0|Q|\eta' \rangle |^2 \over m^2_{\eta'}} = A ,
\label{EQN5.20}
\ee
where $A \equiv A_0(0)$ is the pure Yang--Mills topological susceptibility
in the large--$N_c$ limit.
In the chiral limit of $L$ massless quarks, the topological charge density
$Q(x)$ is directly related to the four--divergence of the axial current
$J_{5, \mu}$, {\it via} the anomaly equation,
$\partial^\mu J_{5, \mu}(x) = 2L Q(x)$, so that
$\langle 0|Q|\eta' \rangle = {1 \over \sqrt{2L}}m^2_{\eta'} F_{\eta'}$,
which can be substituted into Eq. (\ref{EQN5.20}) to give:
\be
A = {m^2_{\eta'} F^2_{\eta'} \over 2L} .
\label{EQN5.22}
\ee
This equation relates the mass $m_{\eta'}$ of the $\eta'$ state, its
decay constant $F_{\eta'}$ and the pure--gauge topological susceptibility $A$.
And in fact Eq. (\ref{EQN5.22}) is verified when putting for $m_{\eta'}$ and
$F_{\eta'}$ their values determined above in Eqs. (\ref{eqn12}) and
(\ref{EQN5.18}).\\
The dominance of the $\eta'$ state in the sum over the meson states at the
right--hand side of Eq. (\ref{EQN5.19}) can also be used to evaluate  the
slope of the topological susceptibility:
\ba
\label{chi'sp}
\chi'_{ch}=\frac{d}{dk^{2}}\chi(k)\Bigg|_{k=0}\simeq
- \frac{|\langle0|Q(0)|\eta'\rangle|^{2}}{m^{4}_{\eta'}}=
-\frac{1}{2L}\,F_{\eta'}^{2}=-\frac{1}{2L}\,(\csti),
\ea
thus recovering the result (\ref{chi'sotto}), derived above from our effective
Lagrangian. This is perfectly natural: using the effective Lagrangian at
tree--level (i.e., using its ``free'' propagators) one gets the results in the
one--hadron pole (i.e., one--hadron intermediate state) approximation.

Summarizing, we have found that the value of $\chi'$ in the chiral limit
$\sup(m_i)\to0$ is shifted from the ``original'' value $-\frac{1}{2L}F_{\pi}^2$
(derived in the absence of an extra $U(1)$ chiral condensate: see Refs.
\cite{spin-crisis}, where $\chi'_{ch}$ is shown to be a relevant
quantity in the discussion of the so--called ``proton--spin crisis'' problem,
and also Ref. \cite{GRTV02}) to the value
$-\frac{1}{2L}F_{\eta'}^2=-\frac{1}{2L}(\csti)$, which also depends on the
quantity $F_X$, proportional to the extra $U(1)$ chiral condensate.\\
Therefore, a measure of this quantity $\chi'_{ch}$, e.g., in lattice gauge
theory, could provide an estimate for the $\eta'$ decay constant $F_{\eta'}$,
and, as a consequence, for $F_X$.
At present, lattice determinations of $\chi'$ only exist for the pure--gauge
theory at $T=0$, with gauge group $SU(2)$ \cite{Briganti-et-al.91} and
$SU(3)$ \cite{EM1992b}.\footnote{It has been recently pointed out in Ref.
\cite{GRTV02} that there can be ambiguities in the definition of $\chi'_{ch}$
in a lattice regularized theory. However, these ambiguities do not apply to
the leading term [of order ${\cal O}(N_c)$] given by Eq. (\ref{chi'sotto}),
but they can affect the next--to--leading term [of order ${\cal O}(N_c^0)$]
in the $1/N_c$ expansion.}

All the above refers to the theory at $T=0$ (or, more generally, for
$T<T_{ch}$).\\
When approaching the chiral transition at $T=T_{ch}$, one expects that
$F_{\pi}$ vanishes, while $F_X$ remains different from zero and the quantity
$\chi'_{ch}$ tends to the value:
\be
\chi'_{ch}\mathop{\longrightarrow}_{T\to T_{ch}}-\unme F_X^2.
\label{chi'Tch}
\ee
Indeed, the quantity $\chi(k)=-i\int d^{4}x\;e^{ikx}\langle{TQ(x)Q(0)}\rangle$
can also be evaluated in the region of temperatures $T_{ch}<T<T_{U(1)}$,
proceeding as for the case $T<T_{ch}$, obtaining the result (already derived in
Ref. \cite{EM1994a}):
\be
\chi(k)=A\frac{k^{2}}{k^{2}-\frac{2A}{F^2_X}},
\ee
in the chiral limit $\sup(m_i)\to0$.\\
Therefore, in the region of temperatures $T_{ch}<T<T_{U(1)}$, $\chi'_{ch}$
is given by:
\be
\label{chi'sopra}
\chi'_{ch}=\frac{d}{dk^{2}}\chi(k)\Bigg|_{k=0}=
-\frac{1}{2}\,F_X^{2},
\ee
consistently with the results (\ref{chi'sotto}) and (\ref{chi'Tch}) found
above: i.e., $\chi'_{ch}$ varies with continuity across $T_{ch}$.
This means that $\chi'_{ch}$ acts as a sort of order parameter for the $U(1)$
axial symmetry above $T_{ch}$: if $\chi'_{ch}$ is different from zero above
$T_{ch}$, this means that the $U(1)$--breaking parameter $F_X$ is different
from zero.

\newsection{Radiative decays of the pseudoscalar mesons}

\noindent
Further information on the quantity $F_X$ (i.e., on the new $U(1)$ chiral
condensate, to which it is related) can be derived from the study of the
radiative decays of the ``light'' pseudoscalar mesons in two photons.
These decays were also studied in Ref. \cite{DiVecchia-Veneziano-et-al.81},
using an effective Lagrangian model, in which only the $q\bar{q}$
chiral condensate was considered.
In this section we want to find the decay rates of the processes
$\pi^0,\eta,\eta',\eta_X\to\gamma\gamma$ (where $\eta_X$ represents the
pseudoscalar meson state, introduced in Section 4, having the same quantum
numbers of the $\eta'$, but a larger mass and a different quark content)
and see which are the effects due to the new $U(1)$ axial condensate, in the
realistic case of $L=3$ light quarks and in the simple case of zero
temperature ($T=0$).\\
To this purpose, we have to introduce the electromagnetic interaction in our
effective model.
First of all, in order to make the Lagrangian invariant under {\it local}
$U(1)$ electromagnetic transformations [$q \to q' = e^{i\theta e \qu}q$, in
terms of the quark fields; the matrix $\qu$ is defined in Eq. (\ref{caricael})
below], we have to replace the derivative of
the field $U$ with the corresponding {\it covariant} derivative $D_\mu U$,
which, by virtue of the transformation property of the field $U$
[$U \to U' = e^{i\theta e \qu} U e^{-i\theta e \qu}$], has the following form:
\be
D_{\mu}U=\partial_{\mu}U+ie A_{\mu}[\qu,U],
\ee
where $A_{\mu}$ is the electromagnetic field (which transforms as:
$A_\mu \to A'_\mu = A_\mu - \partial_\mu \theta$) and $\qu$ is the quark charge
matrix (in units of $e$, the absolute value of the electron charge):
\be
\label{caricael}
\qu=
\left( \begin{array}{ccc}
\frac{2}{3} & \\
& -\frac{1}{3} \\
& & -\frac{1}{3} \\
\end{array}\right).
\ee
Instead,the field $X$ is invariant under a $U(1)$ electromagnetic gauge
transformation and therefore its covariant derivative just coincides with
the ordinary four--derivative: $D_\mu X = \partial_\mu X$.\\
In addition, we have to reproduce the effects of the electromagnetic anomaly,
whose contribution to the four--divergence of the $U(1)$ axial current
($J_{5,\mu}=\bar{q}\gamma_{\mu}\gamma_{5}q$) and of the $SU(3)$ axial currents
($A^{a}_{\mu}=\bar{q}\gamma_{\mu}\gamma_{5}\frac{\tau_{a}}{\sqrt{2}}q$, where
the matrices $\tau_a$, with $a=1,\ldots,8$, are the generators of the algebra
of $SU(3)$ in the fundamental representation, already introduced in the
previous section) is given by:
\be
(\partial^{\mu}J_{5,\mu})^{e.m.}_{anomaly} = 2\Tr(\qu^{2}) G,~~~~
(\partial^{\mu}A^{a}_{\mu})^{e.m.}_{anomaly} =
2\Tr\left( \qu^{2}\frac{\tau_{a}}{\sqrt{2}}\right) G,
\ee
where $G\equiv\frac{e^{2}N_{c}}{32\pi^{2}}\eps F_{\mu\nu}F_{\rho\sigma}$
($F_{\mu\nu}$ being the electromagnetic field--strength tensor), thus
breaking the corresponding chiral symmetries. We observe that
$\Tr(\qu^{2}\tau_{a}) \ne 0$ only for $a=3$ or $a=8$.\\
We must look for an interaction term ${\cal L}_I$ (constructed with the chiral
Lagrangian fields and the electromagnetic operator $G$) which, under a $U(1)$
axial transformation $q \to q' = e^{-i\alpha\gamma_5}q$, transforms as:
\be
U(1)_A:~~{\cal L}_I \to {\cal L}_I + 2\alpha \Tr(\qu^2)G,
\label{prop-u1}
\ee
while, under $SU(3)$ axial transformations of the type $q \to q' = e^{-i\beta
\gamma_5 \tau_a/\sqrt{2}}q$ (with $a = 3,8$), transforms as:
\be
SU(3)_A:~~{\cal L}_I \to {\cal L}_I + 2\beta \Tr\left( \qu^2
{\tau_a \over \sqrt{2}} \right) G.
\label{prop-su3}
\ee
The electromagnetic anomaly term for the field $U$, originally proposed in Ref.
\cite{DiVecchia-Veneziano-et-al.81}, has the following form:
$\unme i G\Tr[\qu^{2}(\ln U-\ln\Ucr)]$.
By virtue of the transformation property of the field $U$ under a $\gru$
chiral transformation ($q_L \to V_L q_L$, $q_R \to V_R q_R$ $\Rightarrow$
$U \to V_L U V_R^\dagger$, where $V_L$ and $V_R$ are arbitrary unitary
matrices \cite{EM1994a,EM2002a}), it is immediate to see that this term
satisfies both the transformation properties (\ref{prop-u1}) and
(\ref{prop-su3}).\\
If we also consider an analogous term for the field $X$, of the form
$\frac{1}{2L}iG\Tr(\qu^{2})(\ln X-\ln\Xcr)$
(where, in our case, $L=3$), then, by virtue of the transformation property
of the field $X$ under a $\gru$ chiral transformation ($X \to \det(V_L)
\det(V_R)^* X$ \cite{EM1994a,EM2002a}), one can see that this term satisfies
the transformation property (\ref{prop-u1}) under a $U(1)$ axial
transformation, while it is invariant under $SU(3)$ axial transformations.
Therefore, if we try to consider a linear combination of the two terms
(inspired by what was done for the $U(1)$ axial anomaly terms in Eq.
(\ref{eqn9}) \cite{EM1994a}), i.e.,
\be
\La_{I}=\unme i\omega_{2} G\Tr[\qu^{2}(\ln U-\ln\Ucr)]
+\frac{1}{2L}i(1-\omega_{2})G\Tr(\qu^{2})(\ln X-\ln\Xcr),
\ee
we immediately see that the property (\ref{prop-u1}) is satisfied for every
value of the parameter $\omega_2$, while the property (\ref{prop-su3}) is
satisfied only for $\omega_2 = 1$.\\
In conclusion, the electromagnetic anomaly interaction term is simply given by:
\be
\label{li}
\La_{I}=\unme iG\Tr[\qu^{2}(\ln U-\ln\Ucr)].
\ee
Therefore, we have to consider the following effective chiral Lagrangian,
which includes the electromagnetic interaction terms described above:
\ba
\label{lem}
\lefteqn{\La(U,\Ucr,X,\Xcr,Q,A^{\mu})=
\frac{1}{2}\Tr(D_{\mu}UD^{\mu}\Ucr)+
\frac{1}{2}\partial_{\mu}X\partial^{\mu}\Xcr }&& \nonumber\\
& & -V(U,\Ucr,X,\Xcr)+\frac{1}{2}iQ\;\omega_{1}\Tr(\ln U-\ln \Ucr) \nonumber\\
& & +\frac{1}{2}iQ(1-\omega_{1})(\ln X-\ln \Xcr)+\frac{1}{2A} Q^{2}
\nonumber\\
& & + \unme iG\Tr[\qu^{2}(\ln U-\ln\Ucr)]
-\frac{1}{4}F_{\mu\nu}F^{\mu\nu},
\ea
where the potential term $V(U,\Ucr,X,\Xcr)$ is the one written in Eq.
(\ref{eqn10}).

We can now describe the predictions of our model about the decay amplitudes and
rates of the processes $\pi^0,\eta,\eta',\eta_X\to\gamma\gamma$.
It is well known that the two final photons in the pseudoscalar--meson decays
are associated with the pseudoscalar operator $\eps F_{\mu\nu}F_{\rho\sigma}$:
so, e.g., the $\pi^0$ decay is reproduced by an interaction of the form
$\pi_3\,\eps F_{\mu\nu}F_{\rho\sigma}$. The covariant derivative does not
contain this term (since the diagonal matrix $\qu$ and the generator $\tau_3$
commute); instead, it is produced by the electromagnetic anomaly interaction
written in Eq.  (\ref{li}) (indeed the chiral symmetry is broken by the
anomaly). The same arguments can be applied  to the other decays. Therefore,
the decay amplitude of the generic process ``$meson\to\gamma\gamma$''
is given by:
\be
A(meson\to\gamma\gamma)=
\langle\gamma\gamma|\La_{I}(0)|meson\rangle.
\ee
Substituting the exponential form (\ref{u,x}) (with $L=3$)
of the field $U$ (we remind the reader that we are considering the case of
zero temperature) into Eq. (\ref{li}), we have the following interaction
Lagrangian:
\be
\label{li1}
\La_{I}=-G\frac{1}{3F_{\pi}} \left( \pi_{3}+\frac{1}{\sqrt{3}}\pi_{8}+
\frac{2\sqrt{2}}{\sqrt{3}}S_{\pi} \right).
\ee
We immediately consider the realistic case of $L=3$ quarks with
masses different from zero. In this case one has to consider the following
quadratic Lagrangian, which can be obtained substituting the exponential
expressions (\ref{u,x}) of the fields $U$ and $X$ into Eq. (\ref{eqn9}):
\ba
\label{l3}
\lefteqn{\La_{2}=\unme\Tr(\partial_{\mu}\Phi\partial^{\mu}\Phi)+
\unme\partial_{\mu}S_{X}\partial^{\mu}S_{X}-\frac{B_{m}}{2F_{\pi}}
\Tr(M\Phi^{2}) }
&&\nonumber\\
&&-c\Big(\frac{1}{F_{\pi}}\Tr\Phi-\frac{1}{F_{X}}S_{X}\Big)^{2}-
A\Big(\frac{\omega_{1}}{F_{\pi}}\Tr\Phi+
\frac{1-\omega_{1}}{F_{x}}S_{X}\Big)^{2},
\ea
where $c\equiv\frac{c_{1}}{\sqrt{2}}\Big(\frac{F_{X}}{\sqrt{2}}\Big)
\Big(\frac{F_{\pi}}{\sqrt{2}}\Big)^3$.\\
Substituting Eq. (\ref{u,x}) (with $L=3$) into Eq. (\ref{l3}), one immediately
sees that the fields $\pi_{1},\pi_{2},\pi_{4},\pi_{5},\pi_{6},\pi_{7}$ are
diagonal, while the fields  $\pi_{3},\pi_{8},S_{\pi},S_{X}$ mix together.
However, neglecting the experimentally small mass difference between the
quarks \emph{up} and \emph{down} (i.e., neglecting the experimentally small
violations of the $SU(2)$ isotopic spin), also $\pi_3$ becomes diagonal and can
be identified with the physical state $\pi^0$. The other physical states can be
found by the diagonalization of the following squared mass matrix, written for
the ensemble of fields $(\pi_{8},S_{\pi},S_{X})$ (originally derived in Ref.
\cite{EM1994c}):
\be
\label{matrice3}
{\cal K}=
\left( \begin{array}{ccc}
\frac{2B}{3}(\tilde{m}+2m_s)&\frac{2B\sqrt{2}}{3}(\tilde{m}-m_s)&0\\
&&\\
\frac{2B\sqrt{2}}{3}(\tilde{m}-m_s)&\frac{6(A\omega_{1}^{2}+c)}{F_{\pi}^{2}}+
\tilde{m}_{0}^{2}&\frac{2\sqrt{3}[A\omega_{1}(1-\omega_{1})-c]}{F_{\pi}F_{X}}\\
&&\\
0&\frac{2\sqrt{3}[A\omega_{1}(1-\omega_{1})-c]}{F_{\pi}F_{X}}&
\frac{2[A(1-\omega_{1})^{2}+c]}{F^{2}_{X}}
\end{array} \right),
\ee
where:
\be
B \equiv \frac{B_{m}}{2F_{\pi}}, ~~~
\tilde{m} \equiv \frac{m_u+m_d}{2}, ~~~
\tilde{m}_{0}^{2} \equiv \frac{2}{3}B(2\tilde{m}+m_s).
\ee
The fields $(\pi_8,S_\pi,S_X)$ can be written in terms of the eigenstates
$\eta$, $\eta'$, $\eta_{X}$ as follows:
\be
\pmatrix{ \pi_8 \cr S_\pi \cr S_X } = \mathbf{C}
\pmatrix{ \eta \cr \eta' \cr \eta_X },
\label{diag}
\ee
where $\mathbf{C}$ is the following $3\times3$ orthogonal matrix:
\be
\label{cambio}
\mathbf{C}=
\left( \begin{array}{ccc}
\alpha_{1} & \alpha_{2} & \alpha_{3}\\
\beta_{1} & \beta_{2} & \beta_{3}\\
\gamma_{1} & \gamma_{2} & \gamma_{3}
\end{array} \right)=
\left(\begin{array}{ccc}
\cos\tilde{\varphi} & -\sin\tilde{\varphi} & 0\\
&&\\
\sin\tilde{\varphi}\,\frac{F_{\pi}}{\sqrt{\cstit}} &
\cos\tilde{\varphi}\,\frac{F_{\pi}}{\sqrt{\cstit}} &
\frac{\sqrt{3}F_{X}}{\sqrt{\cstit}}\\
&&\\
\sin\tilde{\varphi}\,\frac{\sqrt{3}F_{X}}{\sqrt{\cstit}}  &
\cos\tilde{\varphi}\,\frac{\sqrt{3}F_{X}}{\sqrt{\cstit}} &
-\frac{F_{\pi}}{\sqrt{\cstit}}
\end{array} \right).
\ee
Here $\tilde{\varphi}$ is a mixing angle, which can be related to the masses
of the quarks \emph{up, down, strange} (and therefore to the masses of the
octet mesons) by the following relation:
\ba
\label{phitilde}
\tan\tilde{\varphi}=\frac{\sqrt2}{9A}BF_{\pi}\sqrt{\cstit}(m_s-\tilde{m})=
\frac{F_{\pi}\sqrt{\cstit}}{6\sqrt{2}A}(m_{\eta}^{2}-m_{\pi}^{2}),
\ea
where $m^{2}_{\pi}=2B\tilde{m}$ and
$m_{\eta}^{2}=\frac{2}{3}B(\tilde{m}+2m_s)$.
The masses $m_{\eta}$, $m_{\eta'}$, $m_{\eta_{X}}$ are given by the squared
root of the eigenvalues of the squared mass matrix written in Eq.
(\ref{matrice3}) \cite{EM1994c}.

The interaction Lagrangian (\ref{li1}), written in terms of the physical fields
$\pi^0,~\eta,~\eta'$ and $\eta_X$, reads as follows:
\be
\label{limasse}
\La_{I}
\equiv-G\frac{1}{3F_{\pi}}\Big(\pi^{0}+a_{1}\,\eta+a_{2}\,\eta'+
a_{3}\,\eta_{X}\Big),
\ee
where $a_{i}=\frac{1}{\sqrt3}(\alpha_{i}+2\sqrt2\beta_{i})$
$({\rm for}~i=1,2,3)$, so that:
\ba
\label{aibis}
a_{1}&=&\sqrt{\frac{1}{3}}\Bigg(\cos\tilde{\varphi}+
2\sqrt2\sin\tilde{\varphi}\,\frac{F_{\pi}}{{\sqrt{\cstit}}}\Bigg),\\
a_{2}&=&\sqrt{\frac{1}{3}}\Bigg(2\sqrt2\cos\tilde{\varphi}\,
\frac{F_{\pi}}{{\sqrt{\cstit}}}-\sin\tilde{\varphi}\Bigg),\\
a_{3}&=&2\sqrt2\Bigg(\frac{F_{X}}{\sqrt{\cstit}}\Bigg).
\ea
With simple calculations we find the following expressions for the decay
amplitudes:
\ba
\label{api0}
A(\pi^{0}\to\gamma\gamma) & = & \frac{e^{2}N_{c}}{12\pi^{2}F_{\pi}}I,\\
\label{aeta}
A(\eta\to\gamma\gamma)
&=&\frac{e^{2}N_{c}}{12\pi^{2}F_{\pi}}\sqrt{\frac{1}{3}}
\Big(\cos\tilde{\varphi}+
2\sqrt2\sin\tilde{\varphi}\,\frac{F_{\pi}}{{\sqrt{\cstit}}}\Big)I,\\
\label{aeta'}
A(\eta'\to\gamma\gamma)
&=&\frac{e^{2}N_{c}}{12\pi^{2}F_{\pi}}\sqrt{\frac{1}{3}}
\Big(2\sqrt2\cos\tilde{\varphi}\,\frac{F_{\pi}}{{\sqrt{\cstit}}}-
\sin\tilde{\varphi}\Big)I,\\
\label{aetax}
A(\eta_{X}\to\gamma\gamma)
&=&\frac{e^{2}N_{c}}{12\pi^{2}F_{\pi}}
2\sqrt2\Big(\frac{F_{X}}{\sqrt{\cstit}}\Big)I,
\ea
where $I\equiv\varepsilon_{\mu\nu\rho\sigma}
k_{1}^{\mu}\epsilon_{1}^{\nu\ast}k_{2}^{\rho}\epsilon_{2}^{\sigma\ast}$
($k_{1}$, $k_{2}$ being the four--momenta of the two final photons and
$\epsilon_{1}$, $\epsilon_{2}$ their polarizations).
Consequently we derive the following decay rates (in the real case $N_c=3$):
\ba
\label{gammapi0}
\Gamma(\pi^{0}\to\gamma\gamma) & = &
\frac{\alpha^{2}m_{\pi}^{3}}{64\pi^{3}F_{\pi}^{2}},\\
\label{gammaeta}
\Gamma(\eta\to\gamma\gamma)&=&
\frac{\alpha^{2}m_{\eta}^{3}}{192\pi^{3}F_{\pi}^{2}}\Big(\cos\tilde{\varphi}+
2\sqrt2\sin\tilde{\varphi}\,\frac{F_{\pi}}{{\sqrt{\cstit}}}\Big)^{2},\\
\label{gammaeta'}
\Gamma(\eta'\to\gamma\gamma)&=&
\frac{\alpha^{2}m_{\eta'}^{3}}{192\pi^{3}F_{\pi}^{2}}
\Big(2\sqrt2\cos\tilde{\varphi}\,
\frac{F_{\pi}}{{\sqrt{\cstit}}}-\sin\tilde{\varphi}\Big)^{2},\\
\label{gammaetax}
\Gamma(\eta_{X}\to\gamma\gamma)&=&
\frac{\alpha^{2}m_{\eta_{X}}^{3}}{8\pi^{3}F_{\pi}^{2}}
\Big(\frac{F_{X}}{\sqrt{\cstit}}\Big)^{2},
\ea
where $\alpha=e^{2}/4\pi \simeq 1/137$ is the fine--structure constant.

If we put $F_X=0$ (i.e., if we neglect the new $U(1)$ chiral condensate),
the expressions (\ref{aeta})$\div$(\ref{aeta'}) written above reduce to the
corresponding amplitudes derived in Ref. \cite{DiVecchia-Veneziano-et-al.81}
using an effective Lagrangian which includes only the usual $q\bar{q}$ chiral
condensate (so there is no field $\eta_X$!):
\ba
\label{aetadv}
A(\eta\to\gamma\gamma) & = & \frac{e^{2}N_{c}}{12\pi^{2}F_{\pi}}\runte
(\cos\varphi+2\sqrt{2}\sin\varphi)I,\\
\label{aeta'dv}
A(\eta'\to\gamma\gamma) & = & \frac{e^{2}N_{c}}{12\pi^{2}F_{\pi}}\runte
(2\sqrt{2}\cos\varphi-\sin\varphi)I.
\ea
The mixing angle $\varphi$ is now defined as follows
[see Eqs. (\ref{diag})--(\ref{cambio}) with $F_X=0$]:
\ba
\pi_{8}&=&\eta\cos\varphi-\eta'\sin\varphi,\nonumber\\
S_{\pi}&=&\eta\sin\varphi+\eta'\cos\varphi,
\ea
and is related to the quark masses (i.e., to the masses of the octet mesons)
by the following relation [see Eq. (\ref{phitilde}) with $F_X=0$]:
\be
\label{phi}
\tan\varphi\simeq\frac{\sqrt2}{9A}BF_{\pi}^{2}(m_s-\tilde{m})=
\frac{F_{\pi}^{2}}{6\sqrt{2}A}(m_{\eta}^{2}-m_{\pi}^{2}).
\ee
Therefore, the introduction of the new condensate (while leaving the
$\pi^0\to\gamma\gamma$ decay rate unaffected, as it must!) modifies
the decay rates of $\eta$ and $\eta'$ (and, moreover, we also have to consider
the particle $\eta_X$). In particular, it modifies the $\eta'$ decay constant,
already in the chiral limit $\sup(m_i) \to 0$.
Indeed, in this limit, Eq. (\ref{aeta'dv}) [with $\varphi=0$: see
Eq. (\ref{phi})] differs from the corresponding Eq. (\ref{aeta'}) [with
$\tilde{\varphi}=0$: see Eq. (\ref{phitilde})] by the
substitution of the pion decay constant $F_{\pi}$ with the quantity
$F_{\eta'}\equiv\sqrt{\cstit}$, which (as it has been explained in the
previous section for the general case of $L$ light flavours) can be identified
with the $\eta'$ decay constant (in the large--$N_c$ limit).

In conclusion, a study of the radiative decays $\eta\to\gamma\gamma$,
$\eta'\to\gamma\gamma$ and a comparison with the experimental data can
provide us with further information about the parameter $F_X$ and the new
exotic condensate.
For example, from Eqs. (\ref{gammaeta}) and (\ref{gammaeta'}), and using the
experimental values for the various quantities which there appear, i.e.,
\ba
& & F_\pi = 92.4(4) ~{\rm MeV},
\nonumber \\
& & m_\eta = 547.30(12) ~{\rm MeV},
\nonumber \\
& & m_{\eta'} = 957.78(14) ~{\rm MeV},
\nonumber \\
& & \Gamma(\eta\to\gamma\gamma) = 0.46(4) ~{\rm KeV},
\nonumber \\
& & \Gamma(\eta'\to\gamma\gamma) = 4.26(19) ~{\rm KeV},
\ea
we can extract the following values for the quantity $F_X$ and for the
mixing angle $\tilde\varphi$:
\be
F_X = 27(9) ~{\rm MeV},~~~ \tilde\varphi = 16(3)^0.
\ee
The value of $F_X$ is not far from the upper limit
$|F_{X}|\lesssim20\;\rm{MeV}$ obtained from the {\it generalized}
Witten--Veneziano formula derived in Ref. \cite {EM1994c}.
Moreover, the values of $F_X$ and $\tilde\varphi$ so
found are perfectly consistent with the relation (\ref{phitilde}) for the
mixing angle.\footnote{We use for the pure--YM topological susceptibility
the value $A=(180\pm5~\rm{MeV})^{4}$, obtained from lattice simulations
\cite{Teper88,APE90,Alles-et-al.97}.}
We thus see that $F_X$ is small when compared with $F_{\pi}$. In particular,
the corrections due to the presence of the new $U(1)$ condensate are
proportional to the ratio $F_{X}^{2}/F_{\pi}^{2}$ [see Eqs. (\ref{phitilde}),
and (\ref{gammaeta})$\div$(\ref{gammaetax})] and they are of the
order of some $\%$.

\newsection{Conclusions}

\noindent
In this paper we have discussed the role of the $U(1)$ axial symmetry
in QCD both at zero and at finite temperature.
One expects that, above a certain critical temperature $T_{U(1)}$, also the
$U(1)$ axial symmetry will be (effectively) restored. We have tried to see
if this transition has (or has not) anything to do with the usual
$SU(L) \otimes SU(L)$ chiral transition: various possible scenarios
have been discussed in Section 2.
In particular, supported by recent lattice results on the pure--YM topological
susceptibility and the so--called ``chiral susceptibilities'' (which have been
discussed at length in the Introduction), we have analysed a scenario in
which a new $U(1)$--breaking condensate survives across the chiral transition
at $T_{ch}$, staying different from zero up to $T_{U(1)} > T_{ch}$.
This scenario can be consistently reproduced using an effective Lagrangian
model, which also includes the new $U(1)$ chiral condensate: this theoretical
model was originally proposed in Refs. \cite{EM1994a,EM1994b,EM1994c}
and it has been briefly summarized in Sections 3 and 4 for the convenience
of the reader.
This scenario could perhaps be verified in the near future by heavy--ion
experiments, by analysing the pseudoscalar--meson spectrum in the singlet
sector.

In Section 5 (which, together with Section 6, contains the main original
results of this paper) we have analysed the consequences of our theoretical
model on the slope of the topological susceptibility $\chi'$, in the
{\it full} theory with quarks, showing how this quantity is modified by
the presence of the new $U(1)$ chiral order parameter: we have found that
$\chi'$ (in the chiral limit $\sup(m_i) \to 0$) acts as an order parameter
for the $U(1)$ axial symmetry above $T_{ch}$.
This prediction of our model could be tested in the near--future Monte
Carlo simulations on the lattice (but see also Ref. \cite{GRTV02}).

Finally, in Section 6, we have found that the existence of the new $U(1)$
chiral condensate can be directly investigated by studying (at $T=0$) the
radiative decays of the pseudoscalar mesons $\eta$ and $\eta'$ in two photons.
A first comparison of our results with the experimental data has been
performed at the end of Section 6: the results are encouraging, pointing
towards a certain evidence of a non--zero $U(1)$ axial condensate (i.e.,
$F_X \ne 0$). However, one should keep in mind that our results have been
derived from a very simplified model, obtained doing a first--order expansion
in $1/N_c$ and in the quark masses. We expect that such a model can furnish
only qualitative or, at most, ``semi--quantitative'' predictions.
Higher--order terms in $1/N_c$ could give rise to corrections (in the
``real world'' with $N_c=3$) of the same order of magnitude of (or even
larger than!) those induced by the new $U(1)$ axial condensate (having
the form $\sim F_X^2/F_\pi^2$).
Moreover, when going beyond the leading order in $1/N_c$, it becomes
necessary to take into account questions of renormalization--group
behaviour of the various quantities and operators involved in our
theoretical analysis. This issue has been widely discussed in the
literature, both in relation to the analysis of $\chi'_{ch}$,
in the context of the proton--spin crisis problem \cite{spin-crisis},
and also in relation to the study of the $\eta'(\eta)$ radiative decays
\cite{gamma-gamma}.
Further studies are therefore necessary in order to continue this analysis
from a more quantitative point of view.\\
Last, but not least, it would be also very interesting (for a comparison
with future heavy--ion experiments) to extend our present analysis of the
radiative decays to the non--zero--temperature case.
We expect that some progress will be done along this line in the near future.

\vfill\eject

{\renewcommand{\Large}{\normalsize}
}

\vfill\eject

\end{document}